\documentclass[a4paper,12pt]{article}
\usepackage{amssymb,amsmath}
\usepackage{a4wide}
\usepackage{epsfig}

\newcommand{\RR}{{\mathbb{R}}}
\newcommand{\CC}{{\mathbb{C}}}
\newcommand{\ZZ}{{\mathbb{Z}}}
\newcommand{\cA}{{\cal A}}
\newcommand{\cL}{{\cal L}}
\newcommand{\cK}{{\cal K}}
\newcommand{\pa}{\partial}
\newcommand{\ve}{\varepsilon}
\newcommand{\vp}{\varphi}

\newcommand{\sgn}{\mathop{\mathrm{sgn}}\nolimits}

\title{Hopf solitons on compact manifolds}
\author{R.\ S.\ Ward\footnote{email address: richard.ward@durham.ac.uk}
  \bigskip
  \\Department of Mathematical Sciences,
  \\Durham University, Durham DH1 3LE.}
\date{\today}

\begin{document}

\maketitle

\begin{abstract}
\noindent
Hopf solitons in the Skyrme-Faddeev system on $\RR^3$ typically
have a complicated structure, in particular when the Hopf number
$Q$ is large. By contrast, if we work on a compact 3-manifold $M$,
and the energy functional consists only of the Skyrme term
(the strong-coupling limit), then the picture simplifies. There
is a topological lower bound $E\geq Q$ on the energy, and the
local minima of $E$ can look simple even for large $Q$. The aim here
is to describe and investigate some of these solutions, when $M$ is
$S^3$, $T^3$ or $S^2\times S^1$.
In addition, we review the more elementary baby-Skyrme system, with
$M$ being $S^2$ or $T^2$.
\end{abstract}

\section{Introduction}

The Skyrme-Faddeev system \cite{FN97, BS99, MS04} involves maps
$\psi:\RR^3\to S^2$ satisfying a suitable boundary condition, with $\psi$
being characterized topologically by its Hopf number $Q\in\ZZ$.
To get a stable static soliton (hopfion), we need an energy functional
$E[\psi]$ which has the effect of fixing the soliton size.
In the Skyrme-Faddeev case, the energy is $E=E_2+E_4$, where
\begin{equation}  \label{SkyrmeFaddeev}
   E_2=\frac{1}{32\pi^2}\int |d\psi|^2\,d^3x, \quad
    E_4=\frac{1}{128\pi^2}\int |\psi^*\omega|^2\,d^3x.
\end{equation}
Here $\omega$ is the area element on $S^2$.
The effect of the term $E_4$ is to prevent the soliton from shrinking, and the
effect of $E_2$ is to prevent it from expanding. However, a more elementary
way to prevent expansion is simply to take the space $M$ to be compact,
and to use $E=E_4$. In this compact situation, the usual $E_2=\int|d\psi|^2$
term in the energy is not needed, and stable solitons can exist without it:
in a sense the picture is simpler in this \lq strong-coupling limit\rq.
The aim in this note is investigate some of the features of this
simpler system.

So we are dealing with maps $\psi: M\to S^2$ where $M$ is compact,
and the energy functional is $E=E_4$. If $M$ is 3-dimensional, then with
appropriate topological conditions and normalization, one has \cite{SS11}
a topological lower bound $E\geq Q$; more details of this will be given below.
Our main question here is how close $E$ can get to its lower bound, for various
manifolds $M$ and various values of the Hopf charge $Q$. This will be
investigated in what follows, for the three cases $M=S^3$, $M=T^3$
and $M=S^2\times S^1$.

There is another aspect to this story.
The lowest-energy configurations of topological solitons with large topological
charge $Q\gg1$ typically have a rather regular structure. For example,
consider the basic Skyrme model, involving a field $\psi:\RR^3\to SU(2)$,
where the map $\psi$ extended to $\RR^3\cup\{\infty\}$ has degree $Q$.
The static energy, suitably normalized, satisfies the Faddeev bound $E\geq Q$.
For large $Q$, the energy functional has many local minima, but the lowest of
these is believed to resemble a chunk of Skyrme crystal, with $E/Q\approx1.036$
\cite{MS04}. The linear behaviour $E\sim Q$ is compatible with a triply-periodic
lattice-like structure, and that is indeed what one gets.

The case of Hopf solitons in the Skyrme-Faddeev system (\ref{SkyrmeFaddeev})
on $\RR^3$ is, however, very different. Here the dependence of $E$ on $Q$
is sublinear, namely $E\sim Q^{3/4}$. More precisely,
there is a topological lower bound $E\geq CQ^{3/4}$, where
$C=2^{-3/2}\times3^{3/8}$ \cite{VK79}, and it is conjectured that the stronger
bound with $C=1$ holds \cite{W99}. There are many local minima of $E$, and
their energies are consistent with $E \gtrsim Q^{3/4}$ \cite{BS99, HS00, S07}.
The sublinear behaviour of $E$ means that these minimum-energy
configurations on $\RR^3$ cannot resemble chunks of a periodic structure,
and indeed their appearance is typically a tangle of knots and links.
For our basic hopfion system $E=E_4$ on a compact space, however, one has
$E\sim Q$. This raises the possibility that Hopf solitons in this case
might exhibit somewhat more regular large-$Q$ behaviour, at least in
some situations.

If $M$ is 2-dimensional, then the relevant energy bound is $E\geq Q^2$,
where $Q$ is the degree of the map $\psi:M\to S^2$.
We shall begin, in the next section, by reviewing this more elementary
situation, for the two cases $M=S^2$ and $M=T^2$. In both cases, there
are explicit fields which saturate the lower bound: the former has been
noted before, but the latter appears to be new.


\section{Two-Dimensional $M$}

Before dealing with Hopf maps, where $M$ is 3-dimensional, we first
consider the case of maps $\psi:M\to S^2$
where $M$ is a two-dimensional compact Riemannian manifold. Let
$\omega$ denote the area element on $S^2$ with $\int\omega=4\pi$, and
$\eta$ the area element on $M$ with $\int\eta=V$. The normalized energy
$E$ and topological charge (degree) $Q$ of $\psi$ are given by
\begin{equation}  \label{BabySkyrme}
   E = \frac{V}{32\pi^2}\int_M |\psi^*\omega|^2\,\eta,
   \quad Q =  \frac{1}{4\pi}\int_M \psi^*\omega.
\end{equation}
Defining a scalar function $B$ by $\psi^*\omega=B\eta$, and expanding
$\int(B-\lambda)^2\eta\geq0$ where $\lambda$ is a real constant,
immediately gives the lower bound
\begin{equation}  \label{BabyBogBound}
   E \geq Q^2,
\end{equation}
with equality if and only if $B$ takes the constant value $B=\pm4\pi Q/V$. 
In fact \cite{SS11}, every critical point of the functional $E[\psi]$ has
$B$ constant.

The case $M=S^2$ is rather simple, and has been noted before \cite{HK08}.
Here the metric on $M$ is taken to be that of the standard unit 2-sphere.
Let $z\in\CC$ be a stereographic coordinate on the source space $M$,
and $w\in\CC$ a stereographic coordinate on the target space. So the field
is described by a function $w=w(z,\bar{z})$. Then, for any positive integer $Q$,
the bound (\ref{BabyBogBound}) is saturated by
\begin{equation}  \label{S2Soln}
   w(z,\bar{z}) = z^Q / |z|^{Q-1}.
\end{equation}
Thus (\ref{S2Soln}) is a critical point of the functional $E$, in fact a global
minimum in the topological sector labelled by $Q$. Notice that the
corresponding field $\psi$ is
continuous, although it is not smooth if $Q>1$. However, it is smooth on the
complement of the two points $z=0,\infty$, with bounded partial derivatives,
and that is enough for the analysis to work. This rotationally-symmetric
but non-smooth solution has a counterpart for Hopf solitons, as we shall
see below.

Let us turn now to the case of a flat torus $M=T^2$, with the Euclidean
coordinates $(x,y)$ each having period $2\pi$. We use the
unit vector $\phi^j=(\phi^1,\phi^2,\phi^3)$ with $\phi^j\phi^j=1$ to
coordinatize the target sphere $S^2$. Then there is a particularly simple
solution with $Q=2$, which has constant energy density and saturates
the bound (\ref{BabyBogBound}), namely
\begin{equation}  \label{T2Soln}
\phi^1=1-\frac{2}{\pi}|x-\pi|, \quad \phi^2=\sgn(x-\pi) f(x) \cos(y),
   \quad \phi^3=f(x) \sin(y),
\end{equation}
where $f(x)=\sqrt{1-\phi^1(x)^2}$. This is continuous, and smooth except
on the lines $x=0,\pi$.
By contrast, a $Q=1$ solution appears not to exist. In fact, a $Q=1$ field, if
allowed to \lq flow down\rq\ the energy gradient in a numerical simulation,
speads out and approaches
a discontinuous configuration. This is analogous to the situation for
harmonic maps from $T^2$ to $S^2$, where the energy is
$E_2=\int|d\psi|^2\,d^2x$: there also no $Q=1$ solution exists \cite{EW76}.
In this case, however, the field \lq spikes\rq\ rather than spreads out
\cite{CZ97}.


\section{Hopfions on $S^2 \times S^1$}

This section is concerned with the case $M=S^2\times S^1$, and
will describe a highly-symmetric critical point of the functional
$E=E_4$. To begin with, however, we examine the topological lower
bound $E\geq Q$ for general compact $M$. The proof summarized
here is a restatement of the one in \cite{SS11}.

The pullback $F=\psi^*\omega$ is a closed 2-form on $M$, and we
say that $\psi$ is {\em algebraically inessential} if $F$ is exact. If we
represent $\psi$ by the  unit vector field $\phi^j$ as before, then
$\omega=\epsilon_{jkl}\phi^j d\phi^k d\phi^l$.
Let $x^\mu=(x^1,x^2,x^3)$ denote local coordinates on $M$,
and $g_{\mu\nu}$ its metric, with determinant $g$. The energy is
defined to be
\begin{equation}  \label{E4}
   E = \kappa \int |\psi^*\omega|^2\,\sqrt{g}\,d^3x
          = \kappa \int F_{\mu\nu} F^{\mu\nu} \,\sqrt{g}\,d^3x,
\end{equation}
where $F_{\mu\nu}=\epsilon_{jkl}\phi^j (\pa_{\mu}\phi^k)(\pa_{\nu}\phi^l)$,
and where $\kappa$ is some normalization constant.
If we take $\psi$ to be algebraically inessential, then there exists a
1-form $A_\mu$ such that $F_{\mu\nu}=\pa_{\mu}A_\nu-\pa_{\nu}A_\mu$.
Defining
$B^\mu = \frac{1}{2} g^{-1/2} \ve^{\mu\alpha\beta}F_{\alpha\beta}$,
we can write the Hopf number $Q$ as
\[
   Q=\frac{1}{16\pi^2} \int_M B^\mu A_\mu \,d^3x.
\]
Note that $E=2\kappa \int B_{\mu} B^{\mu} \,\sqrt{g}\,d^3x$.
We may take $A_\mu$ to be divergence-free, namely
$\nabla^\mu A_\mu=0$, where $\nabla_\mu$ is the Levi-Civita connection
on $M$. From Stokes's theorem we have
\[
\int_M \nabla^\mu (A^\nu F_{\mu\nu})\,\sqrt{g}\,d^3x =0,
\]
and expanding this gives
\[
  E=2\kappa \int_M A^\mu (\Delta A_\mu) \,\sqrt{g}\,d^3x,
\]
where $\Delta$ is the Hodge-Laplace operator
\[
  \Delta A_\mu = -\nabla^\nu \nabla_\nu A_\mu + R_{\mu\nu} A^\nu.
\]
Now let $\lambda$ be the smallest positive  eigenvalue of $\Delta$
acting on divergence-free 1-forms on $M$. Then we get the bound
\[
  E\geq 2\kappa\lambda \int_M A^\mu A_\mu \,\sqrt{g}\,d^3x.
\]
Combining this inequality with the Cauchy-Schwarz inequality
$||A||.||B||\geq\langle A,B\rangle$ gives
$   E \geq 32\pi^2 \kappa \sqrt{\lambda}\, |Q|$;
and so choosing the normalization factor
$\kappa=1/(32\pi^2\sqrt{\lambda})$ yields
\begin{equation}  \label{HopfBound}
   E \geq |Q|.
\end{equation}
A final point to note is that rescaling the metric of $M$ by a constant
simply rescales $\lambda$ in such a way that the normalized energy $E$
is unchanged. So in each of the examples which follow,
there is no loss of generality in fixing the overall scale of $M$.

We turn now to the specific case $M=S^2\times S^1$. We may fix the
scale by taking the length of the $S^1$ to be $2\pi$, and then the radius
$L$ of the $S^2$ remains a free dimensionless parameter.
To get the appropriate normalization of $E$, we
need the eigenvalue $\lambda$ as described above. In effect, $\lambda$
is the smallest positive  eigenvalue of the Hodge-Laplace operator $\Delta$
acting on divergence-free 1-forms on the $S^2$ with radius $L$,
which \cite{GM75} is $\lambda=2/L^2$, and so we set
$\kappa=L/(32\pi^2\sqrt{2})$. With this normalization, we then
have the bound $E\geq Q$.

Note that this is only valid for algebraically inessential maps $\psi$.
We may view the situation as follows. The $S^1$ factor in $M$ allows the
existence of vortices.  Vortices in the Skyrme-Faddeev system have been
studied previously \cite{HJS04, JH09}, in particular the
evolution of single vortices and bunches of vortices. For a field to be
algebraically inessential, however, we need the net vortex number to be
zero. So in the case of interest here, what we need is an equal
number of vortices and antivortices on $S^2$.

The simplest example of such a configuration is to have a vortex at
one point on the sphere, and an antivortex at the antipodal point. 
Let us use standard spherical coordinates $(\theta,\vp)$ for $S^2$,
and $\chi\in[0,2\pi]$ for $S^1$. This simplest vortex-antivortex field has
the form
\begin{equation}  \label{VAV}
  \phi^j = \left( \sin(f)\cos(\vp\pm\chi), \sin(f)\sin(\vp\pm\chi), \cos(f) \right),
\end{equation}
where the profile function $f=f(\theta)$ satisfies the boundary conditions
$f(0)=0$, $f(\pi/2)=\pi$ and $f(\pi)=2\pi$, and where the upper or lower
sign is chosen according to whether $\theta\in[0,\pi/2]$ or $\theta\in[\pi/2,\pi]$.
This field is continuous, algebraically inessential, and has Hopf number $Q=2$.
It has two rotational symmetries, generated by $\pa_{\vp}$ and $\pa_{\chi}$.
Substituting (\ref{VAV}) into the energy (\ref{E4}) gives a functional
$\widehat{E}[f]$ which is easily minimized numerically, for any given value
of $L$. In particular, we find that the lowest energy is attained when
$L\approx1.51$, and it is $E\approx1.0670\times2$, about
$7\%$ above the topological lower bound. The energy density $|F|^2$ is
peaked at $\theta=0,\pi$, in other words at the location of the vortices.

One may generalize (\ref{VAV}) by replacing $\vp\pm\chi$ with
$m\vp\pm n\chi$ where $m$ and $n$ are integers, but in fact this
gives nothing new. For example if $n>1$, we can re-define $n\chi$
as $\chi$ and rescale the whole space to restore the period of $\chi$
to $2\pi$, thereby effectively rescaling $n$ to unity.
So we get a solution with arbitrary (even) Hopf number $Q$, having
$E_4$-energy $7\%$ above its topological lower bound.
It remains an open question whether this doubly-symmetric
solution is stable under non-symmetric perturbations, and whether it
is minimal-energy in its topological sector. But because its energy
remains close to the lower bound, both of these conjectures would seem
to be plausible.


\section{Hopfions on $T^3$}

This section deals with maps $\psi:T^3\to S^2$, where
$T^3$ is the cubic 3-torus with coordinates $x,y,z$
each having period $2\pi$. So the relevant eigenvalue $\lambda$  equals~1,
and therefore we take $\kappa=1/(32\pi^2)$ in (\ref{E4}) to give the bound
$E\geq|Q|$ for algebraically inessential maps.

This case was investigated using a full 3-dimensional numerical
procedure, brief details of which are as follows. The $xyz$ space is modelled by
an $N^3$ lattice, with the unit vector $\phi^j$ being defined at each
lattice site, and with periodic boundary conditions. The image of a plaquette
(say in $xy$) is a spherical quadrilateral on the target space $S^2$,
and the spherical area of this image represents $F_{xy}$. Then $E=E_4$
is modelled by summing the squares of these areas over all plaquettes,
in all three directions. A conjugate-gradient code then mimimizes $E$.
However, this procedure on its own is rather unstable: a field can easily
become \lq discontinuous\rq\ as it flows down the energy gradient. To avoid
this, one may add an $E_2$ term as in (\ref{SkyrmeFaddeev}), with this
term being multiplied by a parameter $\beta$; and then gradually phasing
out $\beta$ so as to leave the pure case $\beta=0$.

Applying this procedure to an initial $Q=1$ field reveals the same behaviour
as in the $T^2$ case described previously: the hopfion spreads out and
approaches a discontinuous configuration. An initial $Q=2$ field, however,
relaxes to a continuous solution, which is depicted in the left-hand panel
of Figure~1.
\begin{figure}[htb]
\begin{center}
\includegraphics[scale=0.8]{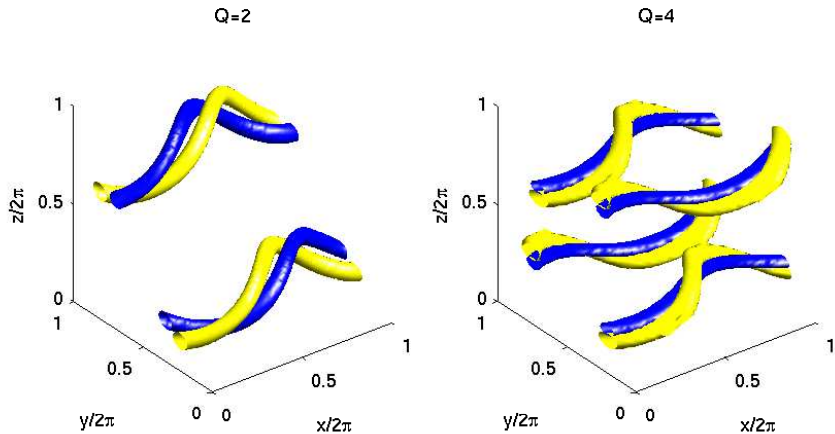}
\caption{$Q=2$ and $Q=4$ solutions on $T^3$ \label{Fig1}}
\end{center}
\end{figure}
The picture shows the curve where $\phi^3=1$, which has two
components (coloured dark); and the curve $\phi^1+\phi^3=\sqrt{2}$, which also
has two components (coloured light). The linking number of $Q=2$ is clear.
The energy density ${\cal E}=\kappa F_{\mu\nu}F^{\mu\nu}$ is almost (but not quite)
constant across $T^3$, and $E/Q\approx1.040$ is $4\%$ above its topological
minimum. The diagram indicates that the solution may be viewed as a
vortex-antivortex pair in the $x$-direction. It has zero net vortex number
in each of the $x$-, $y$- and $z$-directions, as is necessary for it to be 
algebraically inessential.

It is straightforward to see what happens if one changes the periods, by
scaling $x$, $y$ and/or $z$. As was pointed out previously, an overall
scaling has no essential effect. So let us consider allowing the periods
to differ from one another, with the largest period remaining $2\pi$.
Then the bound $E\geq Q$ remains unchanged. Thus suppose the
$x$-period remains $2\pi$,
while the $y$-period becomes $2\pi(1-\ve_y)$ with $\ve_y\geq0$, and the
$z$-period becomes $2\pi(1-\ve_z)$ with $\ve_z\geq0$. The energy
(\ref{E4}) is $E=E_x+E_y+E_z$, where
\[
   E_x = 2\kappa \int(F_{yz})^2\,d^3x,
\]
and similarly for $E_y$ and $E_z$. Then scaling the field to change the
$y$- and $z$-periods changes the energy by
\[
  \delta E = (E_x-E_y+E_z)\ve_y + (E_x-E_z+E_y)\ve_z.
\]
Therefore as long as $E_x>E_y+E_z$ and cyclic, any such scaling will increase
the energy. Now the numerical solution described above has
\[
  (E_x,E_y,E_z)=(0.906,0.587,0.587),
\]
and so indeed satisfies these inequalities. In other words, we cannot lower
the energy by changing the periods.

To obtain higher-charge solutions, it is not enough to simply assemble multiple
copies of the $Q=2$ field described above. For example, doubling in each
of the $x$-, $y$- and $z$-directions produces a configuration with $Q=16$;
but now the periods equal $4\pi$, which changes the normalization factor
$\kappa$, with the result that $E/Q$ doubles. In other words, this
\lq multiple-cell\rq\ field has an energy which is considerably greater
than the topological minimum. Instead, one may begin with (say) a
$Q=4$ initial configuration and allow it to relax numerically, and the result
of such a procedure is depicted in the right-hand plot of Figure~1.
Here we see two vortices and two antivortices, all parallel; the
numerically-minimized energy satisfies $E/Q\approx1.122$.
Once again, it is an open question, but a plausible conjecture, that such
parallel vortex-antivortex fields are the minimal-energy solutions for
each even value of $Q$.


\section{Hopfions on $S^3$}

In this section, we take $M$ to be the standard unit 3-sphere $S^3$.
Every map $\psi:S^3\to S^2$ is algebraically inessential, so
that is not a constraint in this case.
Here we have $\lambda=4$, and so we set $\kappa=1/(64\pi^2)$.
If $Q$ is a perfect square,
then the bound $E\geq Q$ is saturated by an explicit solution which is
invariant under a two-parameter group of symmetries.
If $Q$ is not a perfect square, then this highly-symmetric field has $E>Q$;
but for some values of $Q$ it could still be
the lowest-energy solution. The main aim here is to investigate this.

As just noted, the simplest solutions occur whenever $Q$ is a perfect
square \cite{DF05}; they are analogues of the baby-Skyrme solutions
(\ref{S2Soln}), and satisfy $E=Q$.
They can be described explicitly as follows. Use coordinates
$x^{\mu}=(r,s,t)$, with $r\in[0,\pi/2]$ and $s,t\in[0,2\pi]$, and with the metric
$g_{\mu\nu}$ on $S^3$ being given by
\[
   ds^2 = dr^2 + \cos^2(r)\,ds^2 + \sin^2(r)\,dt^2.
\]
Consider the field given, in terms of the stereographic coordinate $w$, by
\begin{equation}  \label{nsq_field}
   w(r,s,t) = \cot(r)\exp[in(s-t)],
\end{equation}
where $n$ is a positive integer. Then this field has Hopf number $Q=n^2$
and energy $E=n^2$; in fact, the energy density has the constant
value ${\cal E}=n^2/(2\pi^2)$. For $n=1$ this field is simply the
standard Hopf map from $S^3$ to $S^2$. For $n>1$ it is continuous, and
smooth except on the two \lq antipodal\rq\ circles $r=0,\pi/2$ in $S^3$.
It is highly-symmetric, being invariant under the subgroup of the
isometry group of $S^3$ generated by $\pa_s$ and $\pa_t$.

The fact that the field has constant energy density raises the question of how to
visualize it. One way is simply to plot the inverse image $\psi^{-1}(p)$ of a regular
value $p\in S^2$ of $\psi$; this is a curve in $S^3$ possibly having several
components. As an example, the left-hand plot in Figure~2 depicts the
$Q=4$ solution.
\begin{figure}[htb]
\begin{center}
\includegraphics[scale=0.8]{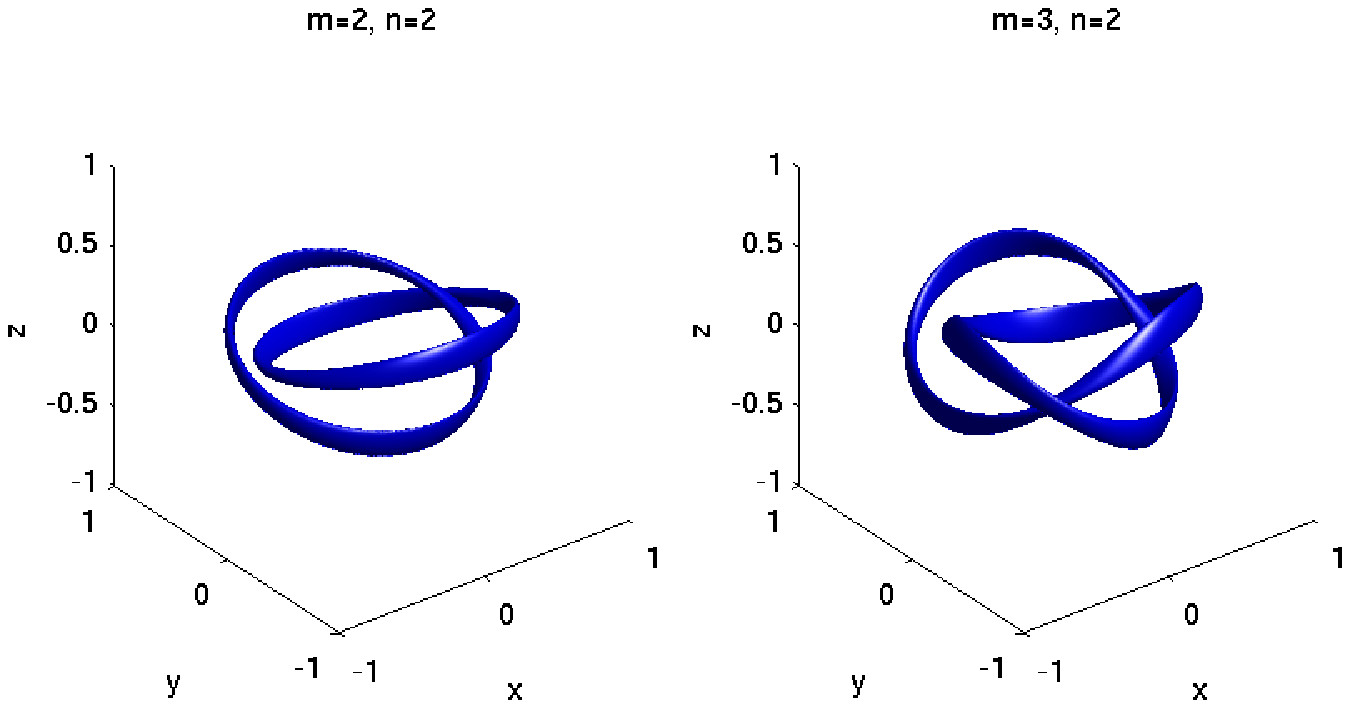}
\caption{The $\cA_{22}$ and the $\cA_{32}$ fields \label{Fig2}}
\end{center}
\end{figure}
We see that the inverse image consists of two linked loops; the inverse
of another regular value would be another linked pair, and the two pairs
would link each other $2\times2=4$ times, as expected.
The $xyz$ space here is a stereographic projection of the original
$S^3$ in which the solution lives. The general $Q=n^2$ case is analogous to
this, with each inverse image $\psi^{-1}(p)$ consisting of $n$ linked loops.

There is an obvious generalization of the solution (\ref{nsq_field})
which remains symmetric under the two rotations, namely
$w(r,s,t) = f(r)\exp(ims-int)$, where $f(r)$ is a suitable profile function,
and $m$ and $n$ are positive integers.
Then the Hopf charge is $Q=mn$. Doubly-symmetric fields of this
type have been studied before \cite{W99}. In the present context, where
the energy functional is just $E_4$, the function $f(r)$
can be determined explicitly \cite{DF05}; the corresponding energy is
\begin{equation}  \label{HopfEmn}
  E_{m,n} = \frac{p-p^{-1}}{2\log{p}} Q,
\end{equation}
where $p=m/n$. Following the notation of \cite{S07}, these
fields are denoted $\cA_{m,n}$. Since $\cA_{m,n}$ and $\cA_{n,m}$ are
essentially the same, we may take $m\geq n$. (This is not true for fields
on $\RR^3$, where $\cA_{m,n}$ and $\cA_{n,m}$ are different, and have
different energies, if $m\neq n$.)
As an example, the energy of the $Q=6$ solution $\cA_{3,2}$
is $E=1.0276\times6$, around $3\%$ above the topological bound.
This field is depicted in the right-hand plot of Figure~2; here, the inverse
image is a single curve, specifically a trefoil knot.

It is worth noting that $w=0$ and $w=\infty$, or equivalently $\phi^j=(0,0,\pm1)$,
are not regular values of the map, and the corresponding inverse images
are single circles in $S^3$ with multiplicity $m$ and $n$ respectively.
If $m\neq n$, then the energy density is not constant, but attains a minimum on
one of these circles and a maximum on the other. In this case,
the energy of $\cA_{m,n}$ is above the topological minimum,
but the excess is small if $m$ is close to~$n$, or equivalently if the number
$p$ in (\ref{HopfEmn}) is close to unity. So one might expect that
if $Q=mn$ with $m\approx n$,
then the minimum-energy field with Hopf charge $Q$ is the highly-symmetric
solution $\cA_{m,n}$. At the other extreme, however (for example if $Q$ is a large
prime), the energy of $\cA_{m,n}$ is relatively high, and the minimum-energy
field is likely to be much less symmetric. What follows gives the results
of investigating this, in a few cases, using a full 3-dimensional
numerical minimization of the energy functional (the appropriate variant
of the one described in the previous section).

Consider first the situation when $m=n+1$, so that $Q=n(n+1)$.
One would certainly expect $\cA_{2,1}$ to be the minimum-energy field
in the $Q=2$ sector, as it is for the Skyrme-Faddeev system on $\RR^3$.
This is indeed borne out by beginning with a non-symmetric deformation of
$\cA_{2,1}$ and observing that it relaxes to $\cA_{2,1}$, with normalized
energy $E=1.0820\times2$.
The next case is $n=2$, so $Q=6$: here the minimum-energy solution
in $\RR^3$ is \cite{BS99, HS00, S07} of the link type $\cL^{1,1}_{2,2}$,
and looks nothing like $\cA_{3,2}$. But for the present system, an initial
configuration of this link type relaxes to $\cA_{3,2}$, in fact the solution
depicted in the right-hand plot of Figure~2. The same thing happens
for several other non-symmetric initial configurations. Consequently,
it seems likely that $\cA_{3,2}$ is indeed the minimal-energy solution
on $S^3$. Finally, this exercise was repeated for the case $n=3$ ($Q=12$),
using various torus-type fields such as $\cK_{3,2}$, $\cK_{4,3}$ and
$\cK_{5,3}$ as initial  configurations (see \cite{S07} for details). Once more,
these relax to the symmetric solution $\cA_{4,3}$, with energy
$E=1.0139\times12$.
For larger values of $n$, the factor in (\ref{HopfEmn}) is even closer to
unity, and so it seems likely that $\cA_{n+1,n}$ is indeed the minimum-energy
field with $Q=n(n+1)$.

Next consider the case $Q=n(n+2)$, in other words $m=n+2$. 
For the Skyrme-Faddeev system on $\RR^3$, the $Q=3$ minimum is
not $\cA_{3,1}$, but rather a \lq buckled\rq\ version $\widetilde{\cA}_{3,1}$.
In our case, however, the numerical results indicate that the symmetry
is maintained, and the minimum-energy $Q=3$ field is $\cA_{3,1}$,
with energy $E=1.2137\times3$. If $n=2$ and $Q=8$, then the $\RR^3$
system prefers non-symmetric fields such as $\cL^{1,1}_{3,3}$
\cite{S07}, but in our case an initial configuration of this type relaxes
to $\cA_{4,2}$. (Note that $\cA_{4,2}$ is closely related to $\cA_{2,1}$,
and has the same value of $E/Q$.) So once again, it is plausible that
$\cA_{n+2,n}$ is the minimum-energy field with $Q=n(n+2)$.

In more \lq extreme\rq\ cases, however, the symmetry is certainly lost.
For example, the field $\cA_{5,1}$ has a relatively high energy, and
as in the $\RR^3$ system it relaxes to a minimum which is non-symmetric
and quite different from $\cA_{5,1}$.


\section{Conclusions}

We have considered local minima of the energy functional
$E[\psi]=\int|\psi^*\omega|^2$ for maps $\psi:M\to S^2$, where $M$
is compact. This is the Skyrme part of the energy in the baby-Skyrme
or Skyrme-Faddeev systems \cite{MS04}, and it is also known as the
$\sigma_2$-energy in the context of differential geometry \cite{ES64}.
In this compact situation, the picture is simpler than when $M=\RR^2$ or
$M=\RR^3$, where an additional energy term such as $E_2=\int|d\psi|^2$
is needed to allow stable solutions. In the 3-dimensional case, the
normalized energy is bounded below by the Hopf number $Q$;
and there are \lq regular\rq\ minima of $E$ even when $Q$ is large.

Several examples were described in this note, but many open questions
remain. For example, in the case $M=T^3$, are the critical points of $E$
necessarily of the parallel vortex-antivortex type, requiring $Q$ to be even?
A more general question is as follows. One may introduce an
additional length scale to the system by adding $\beta E_2$
to the Skyrme energy, where $\beta$ is a constant. As $\beta$ is
increased from zero, we would expect a \lq phase transition\rq\ at some
critical value. This was previously noted \cite{W99} in the simplest
case when $M=S^3$ and $Q=1$, and it would be interesting to study
the phenomenon more generally.


\bigskip\noindent{\bf Acknowledgment.}
The author was supported in part by the UK Particle Science and Technology
Facilities Council, through the Consolidated Grant No.\  ST/J000426/1.


\end{document}